# Magnetic state of Nb(1-7nm)/$Cu_{30}Ni_{70}$ (6nm) superlattices revealed by Polarized Neutron Reflectometry and SQUID magnetometry

Yu. Khaydukov[1,2,7], R. Morari[3,4], D. Lenk[3], V. Zdravkov[3,4], D.G. Merkel[5], B.-K. Seidlhofer[6], C. Müller[3], H.-A. Krug von Nidda[3], T. Keller[1,2], R. Steitz[6], A. Sidorenko[4], S. Horn[3], R. Tidecks[3], and B. Keimer[1]

[1] Max-Planck-Institut für Festkörperforschung, Stuttgart, Germany
[2] Max Planck Society Outstation at the FRM-II, Garching, Germany
[3] Institut für Physik, Universität Augsburg, Augsburg, Germany
[4] D. Ghitsu Institute of Electronic Engineering and Nanotechnologies ASM, Kishinev, Moldova
[5] Wigner Research Centre for Physics, Budapest, Hungary
[6] Helmholtz-Zentrum für Materialien und Energie, Berlin, Germany
[7] Skobeltsyn Institute of Nuclear Physics, Moscow State University, Moscow, Russia

E-mail: y.khaydukov@fkf.mpg.de

**Abstract**. We report results of a magnetic characterization of $[Cu_{30}Ni_{70}(6nm)/Nb(x)]_{20}$ (x=1÷7nm) superlattices using Polarized Neutron Reflectometry (PNR) and SQUID magnetometry. The study has shown that the magnetic moment of the structures growths almost linearly from H = 0 to $H_{sat}$ = 1.3kOe which is an indirect evidence of antiferromagnetic (AF) coupling of the magnetic moments in neighbouring layers. PNR, however, did not detect any in-plane AF coupling. Taking into account the out-of-plane easy axis of the $Cu_{30}Ni_{70}$ layers, this may mean that only the out-of-plane component of the magnetic moments are AF coupled.

## 1. Introduction

Hybrid superconducting/ferromagnet (S/F) heterostructures are intensively studied objects due to their interesting and promising properties [1]. At the moment the main research is concentrated on the study of simple S/F bilayers and S/F/S, F/S/F, and S/F/F trilayers (see [2] and references therein). However, we may expect that both superconducting ([3]-[11]) and magnetic ([12]-[14]) properties of a more complex S/F systems, like $[S/F]_n$ superlattices will differ from those of their constituent elements (S/F bilayers or S/F/S-, F/S/F -trilayers). A difference between the behaviour of the constituent elements and the superlattice is especially expected when the thicknesses of the layers become comparable with the correlation length of superconductivity, $\xi_S$, and magnetism, $\xi_F$, in the respective layers [12]-[15]. In a sense such superlattices can be considered as metamaterials assembled from "atoms" of S/F bilayers.

Our main goal is the design and fabrication of SF metamaterials with unique superconducting and magnetic properties arising from the competition of magnetic and superconducting ordering. As the building blocks of this metamaterial we propose to use Nb/CuNi bilayers investigated in detail before [16]-[19].

There are several reasons for this choice of materials, like very small but non-vanishing solubility of CuNi and Nb (yielding smooth interfaces), good electrical contact and relatively high superconducting correlation lengths $\xi_{S,F} \sim$ 10nm [17]. The aim of this investigation is to search for a possible antiferromagnetic (AF) coupling of $Cu_{30}Ni_{70}$ layers through the Nb spacer. Such an AF coupling can give rise to unique superconducting and ferromagnetic properties of SF metamaterials [13]-[15]. A similar AF coupling has been previously observed by Polarized Neutron Reflectometry (PNR) in Fe/Nb systems [20] for thicknesses of the Nb spacers of $d_{Nb}$ = 1.3nm, 1.7nm, 2.4nm, 2.7nm, 3nm.

## 2. Sample preparation

A series of periodic structures (PS) with nominal composition Si/[Nb($d_{Nb}$)/$Cu_{30}Ni_{70}$(6nm)]$_{20}$/Si ($d_{Nb}$ = 1 ÷ 7nm) were prepared using an automatic magnetron sputtering device UNIVEX at Augsburg University. This machine allows an automatic fabrication of layered structure with specified parameters. All targets are mounted in independent chambers (*target-cells*) and the sample, fixed on the sample holder, is transferred between the target-cells by a robotic manipulator (see Fig. 1).

The typical base pressure in the chambers is $3 \times 10^{-7}$, $6 \times 10^{-7}$ and $5 \times 10^{-6}$ mbar for target-cells, transfer and load-lock chambers, respectively. Pure argon (99.999%, "Linde AG") at a pressure of $8 \times 10^{-3}$ mbar was used as sputter gas. Three targets, Si, Nb and $Cu_{30}Ni_{70}$ (100 mm in diameter), were pre-sputtered for 10 – 15 minutes to remove contaminations before deposition of the samples and 1 minute before each layer deposition. The {111} silicon substrate was etched in pure argon plasma for 5 minutes. After the sample holder was transferred into the cleaned Si-target chamber, the Si-cell was hermetically closed and a 6 nm buffer silicon layer was deposited. The growth rate of the Si film was about 0.4 nm/sec. After deposition the waste gas was rapidly pumped out and the pressure in the target-cell and transfer chamber equalized ($6 \times 10^{-7}$ mbar). Next the sluice was opened and the manipulator picked up the sample holder. Then the sample holder was alternatingly transferred to the Nb target-cell and $Cu_{30}Ni_{70}$ target-cell for deposition of the superlattice structure. To avoid oxidation of the structure, the Si-cap film was grown on the top of the layered structure. The processes of pressure leveling between target-cell and transfer chamber and also sealing/opening of the sluice were repeated for every layer, thus ensuring a clean atmosphere in the transfer chamber during fabrication of the structures. The time spent for gas pumping, pressure leveling process and sample transferring between two target-cells was about 25-30 seconds.

The deposition rates for Nb and $Cu_{30}Ni_{70}$ are 0.4 and 0.5 nm/sec, respectively. The thicknesses of the layers were controlled by the time of the ignited target exposition and checked later using neutron reflectometry. For this purpose, the chambers of the UNIVEX are equipped with a movable shutter between target and sample holder. While pure Nb was sputtered using a pulsed power module, the $Cu_{30}Ni_{70}$ and Si – layers deposition and substrate etching were performed using an AC power module.

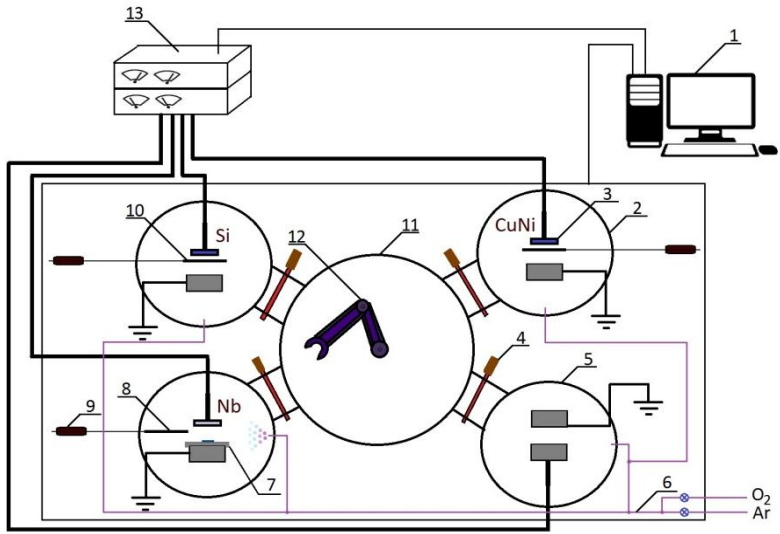

**Figure. 1**. Magnetron sputtering device UNIVEX. 1 - Computer (PC) for controlling the UNIVEX; 2-Target-cell or target chamber; 3 - Target; 4 - Sluice controlled by PC; 5 - Load-lock chamber; 6- Gas pipe; 7 -Sample holder and silicon substrate; 8 - Movable shutter (opened position); 9 - Electric motor for shutter moving; 10 - Movable shutter (closed position); 11 - Transfer chamber; 12 - Robotic manipulator; 13 - AC and Pulsed power modules for magnetron

The structural properties of the layers were characterized using neutron reflectometry. The necessity of using neutron reflectometry instead of the more widely available X-ray reflectometry is due to the extremely low optical X-ray contrast for layers of CuNi and Nb [18],[19]. Figure 2 shows the non-polarized reflectivity curve of the sample with $d_{Nb}$ = 2nm measured at room temperature at the GINA reflectometer [21]. Two peaks at positions $Q_1$ = 0.54 nm$^{-1}$ and $Q_2$ = 1.04 nm$^{-1}$ can be identified and assigned to Bragg reflection from the periodic structure with period $D$. The positions of the peaks are given by $Q_n \approx 2\pi n/D$ ($n$ = 1,2...) and their presence gives evidence for a high structural quality of the superlattice. We were able to reproduce the experimental curve with the scattering length density (SLD) depth profile depicted in the inset to the Fig.1. According to the fit, the nuclear SLD of $Cu_{30}Ni_{70}$ and Nb layers are $\rho_{CuNi}$ = 7.8x10$^{-4}$ nm$^{-2}$ and $\rho_{Nb}$ = 4.3x10$^{-4}$ nm$^{-2}$. The rms roughness of the $Cu_{30}Ni_{70}$/Nb and Nb/$Cu_{30}Ni_{70}$ interfaces is obtained as 0.6 and 1.2 nm, respectively.

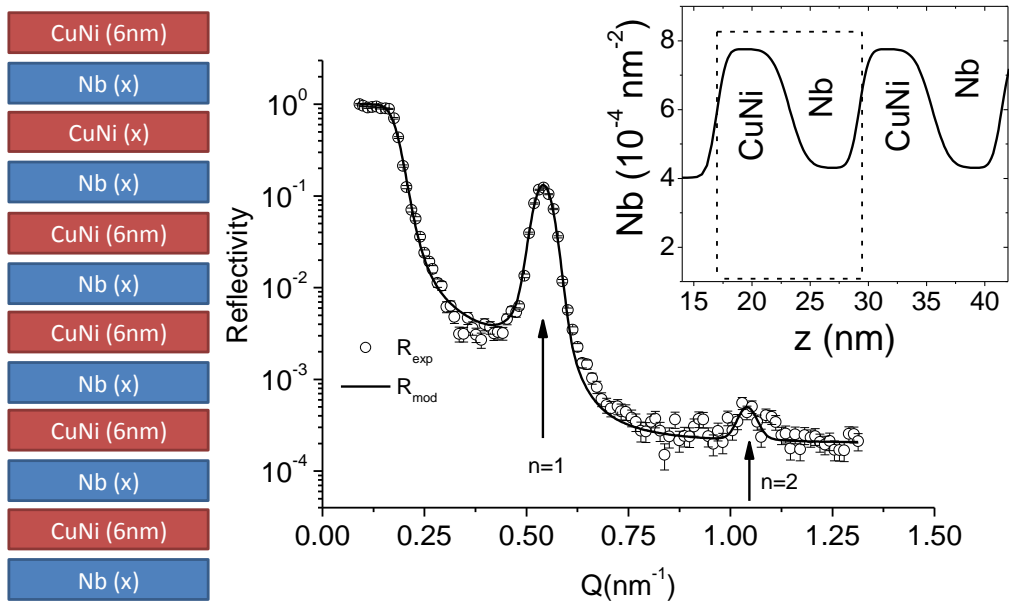

**Figure 2**. Sketch of the structure (left) and room– temperature data measured at the GINA reflectometer. Inset shows part of the SLD depth profile. One "unit cell" of the periodic CuNi/Nb structure is shown by the dashed rectangular.

## 3. Magnetic properties

### 3.1. Single copper-nickel layer

First we have characterized magnetic properties of single $Cu_{30}Ni_{70}$ layers. Figure 3a shows the hysteresis loop of the $Cu_{30}Ni_{70}$(23.5nm) film measured by Superconducting Quantum Interference Device (SQUID) in the Max-Planck Institute for Solid State Research (Stuttgart). The curve is characterized by the coercive field $H_c \approx 150$ Oe and a remanent magnetization which is only 25% of the saturation magnetization. Such a shape of the hysteresis loop is typical for copper-nickel films with the easy axis turned out of plane of the sample [22]. In order to calculate the direction of the easy axis relative to the sample plane, $\beta$, we used the following expression for the magnetic energy of the film

$$E(\alpha) = K_{anys} \sin^2(\beta - \alpha) - M_S H \cos(\alpha), \qquad (1)$$

where $K_{anys}$ - anisotropy energy, $\alpha$, $\beta$ - the angles between the magnetic moment and the easy axis and direction of the external field $H$ (or sample surface, see sketch in Fig. 3a), respectively, $M_S$ - saturation magnetization. The value of $M_S$ can be found from the saturation magnetic moment $m_{sat}$ as $M_S = m_{sat}/(d_{CuNi} \times A)$, where $d_{CuNi}$ and $A$ are the thickness of the $Cu_{30}Ni_{70}$ layer and the sample area, respectively. However, taking into account possible mis-calibration of the film thickness during deposition we decided to use Polarized Neutron Reflectometry (see below) which measures directly the magnetic contrast. Minimizing this energy by varying $\alpha$ for every magnetic field allows to draw a model dependence $m_{sat} \cos(\alpha)$ vs. H (see red curve) and estimate the direction of the easy axis as $\beta \approx 80°$.

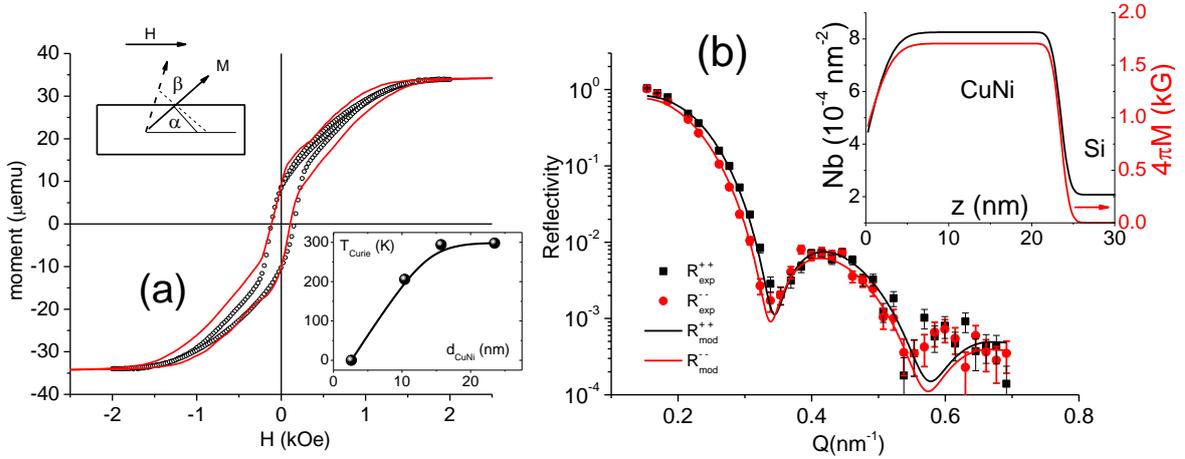

**Figure 3**. (a) Hysteresis loop for the $Cu_{30}Ni_{70}$(23.5nm) layer measured at T = 10K. Inset shows the dependence of the Curie temperature of films on their thickness $Cu_{30}Ni_{70}$(23.5nm). (b) Low temperature PNR curves measured at $T = 150K$ and $H = 4.5kOe$. The error bars if not seen are below the dot size.

The saturation magnetization was measured in a PNR experiment on the same sample. The reflectivity curves $R^+$ and $R^-$ were measured at NREX reflectometer [23] at $T = 150K$ far below the Curie temperature $T_m = 295K$ (see inset in Fig.3a) in a magnetic field $H = 4.5kOe$ (Fig. 3b). Oscillations of the reflectivity curves are caused by the interference over the thickness of the $Cu_{30}Ni_{70}$ film and allowed us also to calibrate the thickness of the $Cu_{30}Ni_{70}$ layer for further preparation of periodic structures. The splitting of $R^+$ and $R^-$ curves is related to the magnetic moment in the film. Fitting the experiment to a model allows us to extract both nuclear SLD of the single film $\rho_{CuNi} = 8.3 \times 10^{-4}$ nm$^{-2}$ and saturation magnetization $4\pi M_{sat} = 1.7kG$. Using the values obtained from PNR for $\rho_{CuNi}$ and $M_{sat}$ we may estimate the magnetic moment per one unit cell atom as

$$m[\mu_B] = M_{sat}[G]/(N\ [\text{cm}^{-3}]\ \mu_B), \qquad (2)$$

where $N$ - is the packing density of $Cu_{30}Ni_{70}$ layer, $\mu_B = 9.27\times10^{-21}$ erg/G - is the Bohr magneton. The density can be derived from the nuclear SLD, which can be written as

$$\rho_{CuNi} = N\times [C_{Ni}\ b_{Ni} + (1 - C_{Ni})\ b_{Cu}], \qquad (3)$$

where $C_{Ni} = 0.7$ -concentration of nickel atoms, $b_{Ni} = 10.3$ fm, $b_{Cu} = 7.7$ fm are neutron scattering lengths for nickel and copper atoms, respectively. Using Eqs. (2) and (3) a magnetic moment of 0.34 $\mu_B$/u.c. or 0.24$\mu_B$/Ni atom can be estimated from experimentally obtained values of $M_{sat}$ and $\rho_{CuNi}$. This value is in agreement with the previously reported 0.27 $\mu_B$/atom for $C_{Ni} = 0.67$ [24].

### 3.2. Periodic structure

The next step is the characterization of the magnetic properties of the periodic structures. Fig. 4a shows the field dependence of the magnetic moment of the sample with $d_{Nb} = 5$nm measured by SQUID magnetometry at a temperature of 13K. The magnetic moment growths almost linearly from zero at $H = 0$ till its saturation value at $H > H_{sat} = 1$kOe. The linear growth of the magnetic moment is the indirect proof of the presence of an antiferromagnetic (AF) coupling of the magnetic moments in the neighbouring layers [24]. Surprisingly, *all* measured samples exhibit similar curves. In order to prove the presence of in-plane AF coupling we have performed low-temperature PNR experiments. The measurements were conducted at the angle dispersive ($\lambda = 0.466$ nm) reflectometer V6 [26]. The protocol of the measurements was as follows. Sample cooled down to $T = 15$K in a magnetic field $H = 5$kOe. At this temperature the field was released and several reflectivity curves were measured at fields $H < H_{sat}$. Fig. 4b shows PNR curves of the sample with $d_{Nb} = 5$nm measured in a magnetic field $H = 400$ Oe at $T = 15$K. Similar to the curve depicted in Fig. 2, the curve is characterized by the presence of the $n = 1$ Bragg peak from the superstructure. Similar to Fig. 3b, Bragg peaks of different polarization are split due to the presence of the magnetic moment in $Cu_{30}Ni_{70}$ layers. Inset to Fig. 4b shows the field dependence of the spin asymmetry of the first Bragg peak $S_1 = (R^+ - R^-)/(R^+ + R^-)$. One can see that this dependence correlates with the field dependence of the magnetic moment depicted in Fig. 4a. A similar behaviour was observed for the sample with $d_{CuNi} = 3$nm.

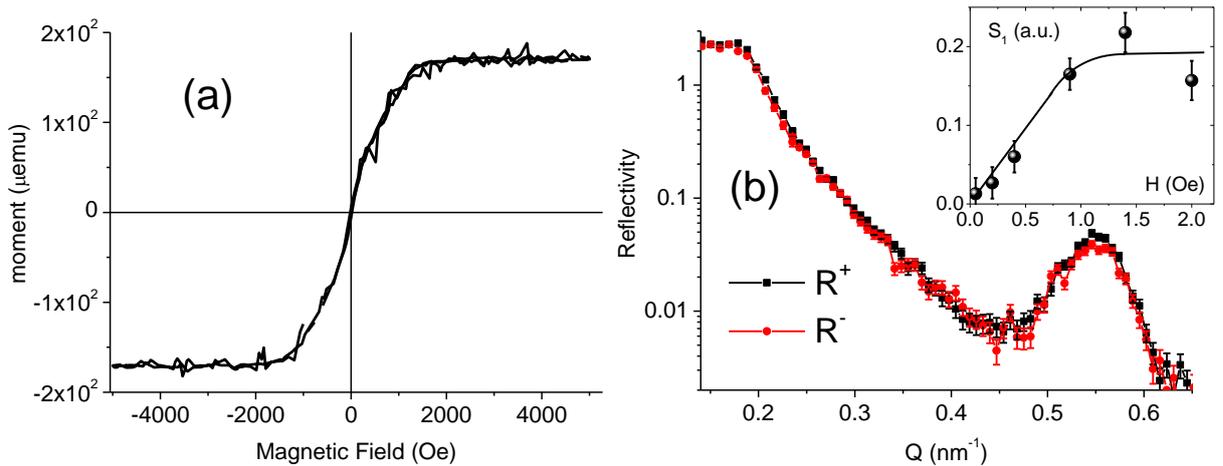

**Figure 4**. (a) Hysteresis loop measured by SQUID magnetometry at $T = 13$K on the sample with $d_{Nb} = 5$nm. (b) PNR curves measured at $T = 15$K in a magnetic field $H = 400$ Oe on the same sample. Dots on the inset show the field dependence of the spin asymmetry of the first Bragg peak. The solid line is shown to guide an eye.

## 4. Discussion and conclusion

Our measurements on single $Cu_{30}Ni_{70}$ films have shown that the magnetic easy axis of the films is out of plane (OOP). This means that the magnetic moments of the $Cu_{30}Ni_{70}$ layers in small magnetic fields will tend to align out of plane. If an AF coupling is presented in the system it would try to align the magnetic moments in the neighbouring layers antiparallel to each other. Such magnetic configuration and its reflectivity curves are shown in Fig. 5a. One can see that in this case there will be the $n = 1/2$ peak due to the doubling of magnetic period in comparison to the structural one. Such peak was indeed seen in PNR experiments with Fe/Nb systems [20]. Another feature is the absence of a splitting at the $n = 1$ peak. We may conclude that this model does not describe our data. Moreover this magnetic configuration leads to a strong increase of the OOP stray field which, in turn, increases the magnetostatic energy. Hence we consider another model, where only out of plane components of the magnetization of neighbouring copper-nickel layers are AF coupled (Fig. 5b). Such a configuration allows decreasing the magnetostatic energy. Since PNR is only sensitive to the in-plane magnetic moment, therefore this configuration will not produce an $n = 1/2$ peak, but will lead to the splitting of curves at the $n = 1$ peak. This model agrees with PNR data depicted in Fig. 4b.

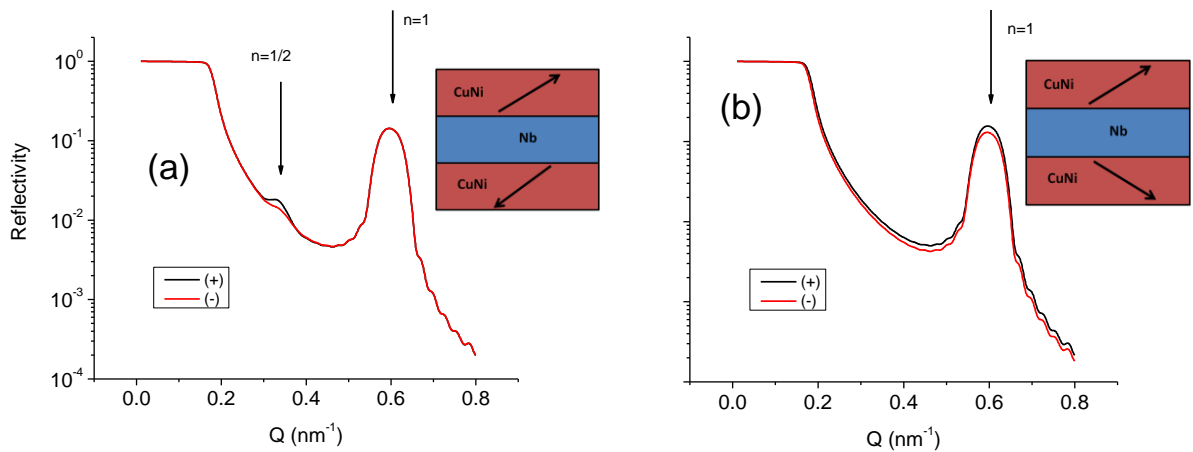

**Figure 5**. Model reflectivity curves for different magnetic configuration depicted in the corresponding insets

In conclusion we report results of the magnetic characterization of $[Nb(d_{Nb})]/Cu_{30}Ni_{70}(6nm)]_{20}$ ($d_{Nb}$ =1÷7nm) superlattices using SQUID magnetometry and polarized neutron reflectometry. The study has shown that in all structures the magnetic moment growths almost linearly from $H = 0$ to $H_{sat}$ = 1.3kOe which is an indirect evidence of antiferromagnetic coupling of moments in neighbouring layers. PNR however did not detect any AF coupling in-plane. Taking into account an out-of-plane easy axis of the $Cu_{30}Ni_{70}$ layers, this probably indicates that only the out of plane component of magnetic moments is AF coupled. Further depth and out-of-plane sensitive techniques like X-ray magnetic scattering are required to answer this question explicitly.

The authors would like to thank A. Petrzhik for assistance in measurement of magnetic moment of single $Cu_{30}Ni_{70}$ layers and T. Kraus and W. Reiber for their help in preparation of the samples. This work is based upon experiments performed at the NREX instrument operated by Max-Planck Society at the Heinz Maier-Leibnitz Zentrum (MLZ), Garching, Germany and partially supported by the DFG collaborative research center TRR and DFG grant No HO 955/9-1.

## References
[1] Buzdin A 2005 *Rev. Mod. Phys.* **77** 935
[2] Lenk D *et al*. 2016 *Beilstein J. Nanotechnol*. **7** 957
[3] Chien C L and Reich D H 1999 *J. Magn. Magn. Mater.* **200** 83

[4]  Verbanck G, Potter C, Schad R, Beilen P, Moshchalkov V and Bruynseraede Y 1994 *Physica C* **235–240** 3295
[5]  Koorevaar P, Coehoorn R and Aarts J 1995 *Physica C* 248 61
[6]  Mattson J, Potter C, Conover M, Sowers C and Bader S 1997 *Phys. Rev.* B **55** 70
[7]  Verbanck G, Potter C, Metlushko V, Schad R, Moshchalkov V and Bruynseraede Y 1998 *Phys. Rev.* B **57** 6029
[8]  Prischepa S, Cirillo C, Bell C, Kushnir V, Aarts J, Attanasio C and Kupriyanov M 2008 *JETP Lett.* **88** 431
[9]  Armenio A A, Cirillo C, Iannone G, Prischepa S and Attanasio C 2007 *Phys. Rev.* B **76** 024515
[10] Huang S, Liang J-J, Tsai T, Lin L, Lin M, Hsu S and Lee S 2008 *J. Appl. Phys.* **103** 07C704
[11] Cirillo C, Bell L C, Iannone G, Prischepa S, Aarts J and Attanasio C 2009 *Phys. Rev.* B **80** 094510
[12] Sá de Melo C A R. 2000 *Phys. Rev.* B **62** 12303
[13] Proshin Y, Izyumov Y and Khusainov M 2001 *Phys. Rev.* B **64** 064522
[14] Halterman K and Valls O 2004 *Phys. Rev.* B **69** 014517
[15] Bakurskiy S, Kupriyanov M, Baranov A, Golubov A, Klenov N,. Soloviev I 2015 *JETP Lett.* **102** 586-593
[16] Zdravkov V, Sidorenko A, Obermeier G, Gsell S, Schreck M, Müller C, Horn S, Tidecks R and Tagirov L 2006 *Phys Rev. Lett.* **97** 057004
[17] Zdravkov V *et al*. 2010 *Phys. Rev.* B **82** 054517
[18] Khaydukov Y *et al*. 2015 *J Supercond Nov Magn* **28** 1143
[19] Khaydukov Y, Morari R, Soltwedel O, Keller T, Christiani G, Logvenov G, Kupriyanov M, Sidorenko A and Keimer B 2015 *J. of Appl. Phys.* **118** 213905
[20] Rehm C, Nagengast D, Klose F, Maletta H and Weidinger A 1997 *Europhys. Lett*. **38** 61-66
[21] Bottyán L, Merkel D, Nagy B, Füzi J, Sajti S, Deák L, Endrőczi G, Petrenko A and Major J 2013 *Rev. Sci. Instrum.* **84** 015112
[22] Ruotolo A, Bell C, Leung C and Blamire M 2004 *J. Appl. Phys.* **96** 512
[23] http://www.mlz-garching.de/nrex
[24] Rusanov A, Boogaard R, Hesselberth M, Sellier H and Aarts J 2002 *Physica* C **369** 330
[25] Granberg P, Isberg P, Svedberg E, Hjörvarsson B, Nordblad P and Wäppling R 1998 *J. Magn. Magn. Mater.* **186** 154
[26] Paul A, Krist T, Teichert A and Steitz R 2011 *Physica* B **406** 1598–1606